\documentclass{article}
\usepackage[utf8]{inputenc}
\usepackage{cite}
\usepackage{amsmath,amssymb,amsfonts}
\usepackage{algorithmic}
\usepackage{graphicx}
\usepackage{graphicx,color}
\usepackage{textcomp}
\usepackage{amssymb}
\usepackage[utf8]{inputenc}
\usepackage{float}
\usepackage{amsmath}
\usepackage{hyperref}
\usepackage{wrapfig}
\usepackage{authblk}

\def\BibTeX{{\rm B\kern-.05em{\sc i\kern-.025em b}\kern-.08em
    T\kern-.1667em\lower.7ex\hbox{E}\kern-.125emX}}
\begin{document}
\title{Power law dynamics in genealogical graphs\\
}

\author[1]{Francisco Leonardo Bezerra Martins}
\author[1]{José Cláudio do Nascimento}
\affil[1]{\small \textit{Federal University of Ceará, Teleinformatics Engineering Department, Pici Campus, Block 725, 60455-970, Fortaleza, Ceará, Brazil}}

\maketitle 

\begin{abstract}
Several populational networks present complex topologies when implemented in evolutionary algorithms. A common feature of these topologies is the emergence of a power law. Power law behavior with different scaling factors can also be observed in genealogical networks, but we still can not satisfactorily describe its dynamics or its relation to population evolution over time. In this paper, we use an algorithm to measure the impact of individuals in several numerical populations and study its dynamics of evolution through nonextensive statistics. Like this, we show evidence that the observed emergence of power law has a dynamic behavior over time. This dynamic development can be described using a family of $q$-exponential distributions whose parameters are time-dependent and follow a specific pattern. We also show evidence that elitism significantly influences the power law scaling factors observed. These results imply that the different power law shapes and deviations observed in genealogical networks are static images of a time-dependent dynamic development that can be satisfactorily described using $q$-exponential distributions.
\end{abstract}

 \begin{center}
\end{center}

\section{Introduction}
A common property of many populational networks is that the degree of connectivity between individuals follows a power law distribution \cite{barabasi1999emergence, giacobini2005takeover, payne2007takeover, payne2008influence, payne2009evolutionary, broido2019scale, reed2002gene}. Experimental results on the dynamics of populational structures have been presented \cite{alba2005exploration, whitacre2008self, clauset2009power}, and a technique that encodes a genealogical history tracing the genetic flow and impact of individuals has revealed the power law with different shapes (or scaling factors) \cite{whitacre2009making}.

Several different forms of probability distributions observed in empirical data have been successfully described by
Nonextensive Statistical Mechanics \cite{abe2001nonextensive, gell2004nonextensive, tsallis2009introduction}, whose formalism is derived from a proposed generalization of the Boltzmann-Gibbs entropy \cite{tsallis1988possible}. Among these, the $q$-exponential distribution family \cite{picoli2009q} is given by
\begin{equation}
p_{qe}(x)=p_0\left[1-(1-q)\frac{x}{x_0}\right]^{\frac{1}{1-q}},
\label{eq1}
\end{equation}
where $x_0$ and $q$ are variable parameters and the inequality $1-(1-q)x/x_0 \geq 0$ and the normalization condition, $p_0=(2-q)/x_0$, must be satisfied. For $q<1$, $p_{qe}$ has finite value for any finite real $x$ since, by definition, $p_{qe} (x)=0$ for $1-(1-q)x/x_0 <0$. For $q>1$, $p_{qe}$ exhibits an asymptotic behavior based on the power law,
\begin{equation}
p_{qe}(x) \sim x^{-\frac{1}{q-1}},
\end{equation}
where the $q$-exponential distribution corresponds to a Burr-type distribution \cite{burr1942cumulative} and to the Zipf-Mandelbrot’s law \cite{mandelbrot1982fractal}. The $q$-exponential distribution is, therefore, a generalization of these distributions, able to represent both the ``heavy tailed'' and the ``light tailed'' distributions. Confirmation of this generalization has been corroborated by several empirical works that addressed diverse subjects, such as the distribution of the population in cities \cite{malacarne2001q}, surnames \cite{yamada2008q}, human behavior \cite{takahashi2011depressive, takahashi2008psychophysics, cajueiro2006note}, circulation of magazines \cite{picoli2005statistical}, delay of trains \cite{briggs2007modelling}, financial markets \cite{politi2008fitting, jiang2008scaling, kaizoji2006interacting, kaizoji2004inflation}, citations in scientific articles \cite{anastasiadis2009characterization, tsallis2000citations}, and even DNA sequences \cite{oikonomou2008nonextensive}.

The dynamics of complex populational networks need a theoretical basis to interpret experimental results \cite{alba2005exploration, whitacre2009making, whitacre2008self} and to understand and solve many open problems \cite{payne2013complex}. Studying the models and conjectures around these problems, as in \cite{cipriani2019dynamical}, can help us understand the evolution of social networks \cite{topirceanu2018weighted}, bitcoin networks \cite{holtz2013evolutionary}, and wealth distribution \cite{kondor2014rich}. In this paper, we study the different power law behaving distributions observed in several numerically generated genealogical networks and describe its dynamics over time using Tsallis’s $q$-exponential distribution. Ultimately, we show that the different power law shapes and deviations observed are a product of a time-dependent dynamic evolution that can be described by the $q$-exponential distribution family.

\section{Methodology}

\label{Methodology}
Here, we describe the methodologies necessary to carry out the processes and numerical experiments presented. These include:
(1) The creation of genealogical trees in the form of graphs from the genetic and historical information of each individual in several populations;
(2) The quantification of the impact of each individual within the family trees;
(3) The creation of the correspondent individual impact value probability density distributions;
(4) And the determination of the best fitting $q$-exponential curves for each probability density distribution.

\subsection{Event Takeover Value (ETV) algorithm}
\label{ETV}
In population-based optimization algorithms, a good measure of the impact of an individual on the dynamics of the entire
population can be obtained by analyzing how said individual, and its offspring, performed in survival and reproduction over the generations.
The Event Takeover Value (ETV) algorithm, proposed in \cite{whitacre2009making}, is an algorithm that measures the impact of an individual on the population dynamics through genealogical graphs. Said algorithm is also useful for observing how the genetic material of an individual is able to spread through the future populations.

\subsubsection{ETV calculation procedure}
Let us consider that, in each generation, $N$ individuals are created. We denote the $i$-th birth by $i$ for $1\leq i \leq N$, and each new generation by $j$ for $1 \leq j \leq t$, where $t$ is the number of generations at the end of each simulation. Like this, any individual in the family tree can be represented by $(i,j)$, and it is possible to count the number of individuals in the population who are historically connected to any ancestor in any generation. The number of individuals historically connected to the $(i,j)$ ancestor, after $k$ generations subsequent to $j$ (where $j \leq k \leq t$), represents the impact of this individual on the $j+k$ generation, denoted by $\iota=ETVgen(i,j,k)$. In the genealogical graph, $\iota$ also represents the number of links from the $(i,j)$ individual to individuals descended from this one in the $j+k$ generation. So, the maximum dissemination power of the genetic material, for any $(i,j)$ individual, is the maximum $ETVgen$ for all $t-j$ subsequent generations after its creation,
\begin{equation}
ETV_{i,j}(t)=max\{ETVgen(i,j,k)\}_{k=j+1}^{t}.
\end{equation}

By definition, note that: (1) The $ETV_{i,j}(t)$ value is always greater than or equal to 1, and is limited by the number of individuals in the population in each generation. Also, since we consider that every generation has $N$ individuals\footnote{In cases where the generations do not have a fixed number of individuals, the $ ETV_{i,j}(t)$ value will be limited to the number of individuals in the generation with the larger population.}, $1\leq ETV_{i,j}(t)\leq N$; (2) At each new generation, the $ ETV_{i,j}(t)$ value is updated, therefore it is dependent on the number of generations at the end of each simulation, $t$. This characteristic is essential to observe the power law dynamics thereof.

\subsubsection{ETV frequency measurement}
If we denote by $n(x)$ the number of individuals with $ETV_{i,j}(t)=x$ and $\cal N$ the number of individuals in the genealogical network, then the frequency of the $x$ value, $n(x)/{\cal N }$, approximates the probability of  $ETV_{i,j}(t)=x$,
\begin{equation}
\text{Pr}[ETV_{i,j}(t)=x]\sim\frac{n(x)}{\cal N }.
\end{equation}

If the number of individuals in each generation\footnote{If the number of individuals generated in each generation is not fixed, then ${\cal N}=\sum_{j=1}^{N} N_j$, where $N_j$ is the number of individuals created in the $j$-th generation.} is $N$, then ${\cal N}=Nt$. To increase the reliability on the $n(x)/{\cal N }$ frequency, we must use the law of large numbers. Like this, we can run the algorithm $R$ times and calculate a more reliable frequency to approximate the probability,
\begin{equation}
 \text{Pr}[ETV_{i,j}(t)=x] \approx \frac{\sum_{r=1}^{R} n_r(x)}{\sum_{r=1}^{R} {\cal N}_r },
 \label{freq}
\end{equation}
where $n_r(x)$ is the number of individuals with $ETV_{i,j}(t)=x$ and ${\cal N}_r$ is the number of individuals in the $r$-th family tree. In our simulations we use $R=20$.

\subsubsection{Genetic hitchhiking}
The genetic hitchhiking, effect in which the impact of an individual on a population is shared with its ancestors, was here disregarded since it may not reflect the reality of many phenomena found in populational networks. For example, under the ETV metric, Genghis Khan was a very high-impact individual in the Asian population \cite{zerjal2003genetic, derenko2007distribution}. If we consider the effect of genetic hitchhiking, his parents, grandparents, and other ancestors would also have the same impact. This approach can be useful in other contexts, where only the individual impact is not enough and the weight of genetic contribution must be shared by the individuals' ancestors. However, in many populational networks, the ascendants' contribution is limited only to generating the individual with the greatest impact, not causing significant direct contributions in the generations in which they were alive. Thus, in our numerical experiments, genetic hitchhiking is disregarded through an ancestrality detachment mechanism\footnote{For more details on the genetic hitchhiking effect, and how to eliminate it, the reader should consult \cite{whitacre2009making}.}.

\subsection{Genetic algorithm}
\label{GA}

In \cite{whitacre2009making}, the authors conducted experiments using evolutionary algorithms to analyze the resulting ETV distributions for several well-known test problems. In their results, it was found that genealogical networks are little sensitive to the number of individuals in a population or the ﬁtness landscape on which evolution occurs. On the other hand, the population updating strategy and several conversion delaying mechanisms were found to contribute to the observation of power law deviations in the population dynamics. Especially so when combined.

In order to cover the conditions necessary for the formation of the power law and its more severe deviations \cite{whitacre2009making,albert2002statistical}, we employ similar experiments to those shown in \cite{whitacre2009making} while using different test problems, different population updating strategies, and different conversion delaying mechanisms.
More specifically, we implemented a genetic algorithm (GA) with the following settings:
\begin{itemize}
  \item Traveling salesman problems as main test problems.
  \item Fixed population of 100 individuals.
  \item Roulette selection.
  \item Crossover by direct analysis of the ``genetic code'' of the involved individuals.
  \item Severe mutation mechanics.
  \item Analysis with the use or disuse of elitism.
  \item Insertion of historically uncoupled individuals.
  \item Implementation of aging and maximum edge limitation.
\end{itemize}

\subsubsection{Traveling salesman problems as main test problems}
The Traveling Salesman Problem (TSP) is a combinatorial optimization problem based on determining the shortest distance circuit to go through a number of points, passing only once at each point and finishing the circuit at the starting point. It is one of the most famous problems in computational mathematics that has numerous direct applications but still lacks an effective solution method for the general case.

Of particular interest to us is the fact that, in a TSP, individuals are represented by its route (a numerical combination of points), which, given a big enough number of points, can act as a ``genetic code'' in our analyzes. This allows us to track when and which ``traits'' of an individual are being spread through the future populations in the genealogical graphs. Further, the TSP presents several characteristics of power law behaving complex systems such as unclear or multiple optimal solutions, unknown complexity, and non-linear dynamics.

\subsubsection{Fixed population of 100 individuals}
In \cite{whitacre2009making}, the authors show that genealogical network dynamics are little sensitive to the number of individuals in a population. Knowing this, we work with a fixed population of 100 individuals per generation. In each new generation, 100 new individuals are ``born'' and may take the place of the individuals of the previous generation depending on the use of elitism and its fitness value.

\subsubsection{Roulette selection}
The selection of individuals for reproduction (crossover) is realized through roulette selection, a method in which individuals with higher fitness are more likely to be selected. This selection method reproduces reasonably well the concept of preferential attachment in its most general form\footnote{Concept in which new network nodes (individuals) are more likely to interact with the nodes with the higher annexation probabilities (usually of higher fitness). This effect is present in several real complex systems, and it plays an important role in the development of free-scale networks with power law structures \cite{barabasi1999emergence, broido2019scale}.} \cite{newman2003structure, albert2002statistical}.

\subsubsection{Crossover by direct analysis of the ``genetic code'' of the involved individuals}
The crossover method employed was the Enhanced Edge Recombination (EER) method\footnote{The EER, while more time-consuming and of more difficult implementation, is generally superior when compared to the Order 1, Order Multiple, and PMX crossover methods. Detailed descriptions on the operation and implementation of the EER method are available in \cite{merz2002comparison, larranaga1999genetic}.}, a TSP crossover method that uses information from the genetic links of the parents in the creation of the offspring. We chose this method due both to the easiness in determining the dominant parent (necessary for the calculation of $ETVgen$), through direct analysis of the genetic links of the parents and their offspring, as well as the intrinsic characteristic of the method in reducing the amount of ``implicit mutations'' during the crossover process. The smaller the number of ``implicit mutations'', the greater is our control over the number of historically uncoupled individuals, and the more accurate will be our analyzes of the population dynamics when considering these cases. The EER method was employed with a probability of 0.9.

\subsubsection{Severe mutation mechanics}
The mutation method chosen was the sub-string inversion mutation, where a string of 2 or more genes is randomly chosen and inverted in the individual's code, here employed with a probability of 0.05. This method was chosen due to its great variability capacity and additional algorithm conversion delay.

\subsubsection{Analysis with the use or disuse of elitism}
Several evolutionary algorithms used in research papers focused on power law behaviors use elitism as default. Elitism, or elitist selection, is a population updating mechanic that guarantees that the population fitness in the GA will not decrease from one generation to the next\footnote{This is achieved by comparing the fitness values of the parents and their offspring after reproduction. If the fitness value of the offspring is lower than that of the parent, then the offspring is discarded, and the parent takes its place in the next generation.}. Elitism is a  mechanic that, while greatly increasing the conversion speed, also inevitably causes power law deviations since, at minimum, a significant number of low-impact individuals are prevented from existing. We will analyze the influence of elitism's use (or disuse) in the populations and power law dynamics.

\subsubsection{Insertion of historically uncoupled individuals}
In order to cause even more ``disturbance'' and cover the cases in which the power law deviates the most \cite{whitacre2009making}, we implemented two mechanisms for the insertion of historically uncoupled individuals. (1) The first occurs in the genetic hitchhiking elimination process, in which some individuals are detached from their ancestors and then considered as ``new'', history-less individuals. (2) The second mechanism occurs using a custom method. This method consists of analyzing, at each generation, the number of times in which each new individual and its reverse\footnote{Keep in mind that here an individual is a route. As an example, for a 4-point TSP, 1-3-2-4-1 is a valid individual, and its reverse is 1-4-2-3-1.} are repeated. If an individual is present in the population more than once, but its reverse is not present, one of these repeated individuals is  replaced by its reverse (which has different coding but the same fitness value). So, whenever an individual of a different genetic code is acquired and repeated, its reverse is created and inserted into the population.

In this way, we guarantee that there will always be two or more different ``genes'' dominating the population. This process increases diversity and decentralizes the dominance of the fittest individuals. The increased diversity occurs simply because different individuals are being inserted into the population (genetically/historically uncoupled individuals). As for the dominance decentralization, the fact that these individuals are different, but have the same fitness as their original reverse, causes a trend of dominance by these individuals to arise in distinct portions of the population. So, new individuals with the same fitness acquire a portion of the population to dominate in the next generations.

\subsubsection{Implementation of aging and maximum edge limitation}
According to \cite{albert2002statistical}, aging and maximum edge limitation are mechanics that cause deviations from the power law behavior. For the aging mechanism, we set a maximum age limit, $m$, up to which each individual may stay alive. An individual dies by being replaced by its offspring. For the maximum edge limitation mechanism, we limit the number of links between individuals in the genealogical networks, i.e., we limit the maximum $ETVgen$ an individual might attain. \bigskip

With these settings, we can implement a genetic algorithm (GA) that solves several symmetrical TSPs while storing the necessary information to construct genealogical trees. These genealogical trees will be evaluated using the ETV algorithm (section \ref{ETV}), and the resulting impact value probability distributions can be analyzed.

\subsection{$q$-exponential fitting method}
\label{Dfm}
Investigating the power law evolutionary dynamics in the resulting $\text{Pr}[ETV_{i,j}(t)=x]$ distributions (or $ETV_{i,j}(t)$ probability distributions) will require us to determine the best-fitting $p_{qe}(x)$ curves across all $t$ generations. In this section, we demonstrate the method employed to calculate the values of the $q$, $x_0$, and $p_0$ parameters for each $p_{qe}(x)$ curve. For simplicity of notation, we will represent $\text{Pr}[ETV_{i,j}(t)=x]$ by PETV, $ETV_{i,j}(t)$ by ETV, and $p_{qe}(x)$ by $p_{qe}$.

\begin{figure}[!ht]
\centering
\includegraphics[scale=0.55]{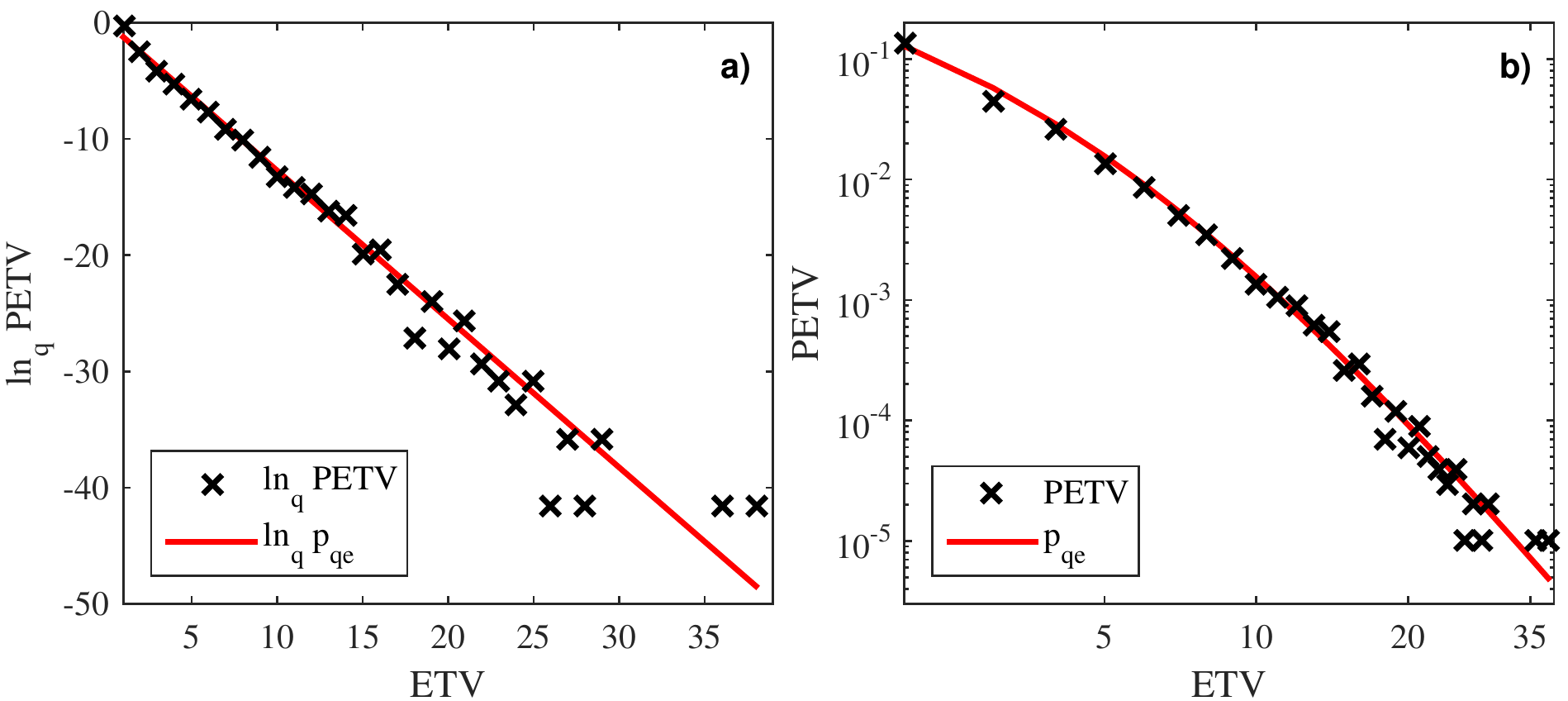}
\caption{\textbf{$\times$ markers:} mean values of a) $ln_{q}$PETV vs. ETV, and b) PETV vs. ETV, in 20 simulations. \textbf{Red curves:} a) resulting $ln_{q}\; p_{qe}$ curve (exhibited in $q$-logarithm scale), and b) resulting $p_{qe}$ curve (exhibited in log-log scale), for $q=1.19$, $p_0=1.0421$, and $x_0=0.7773$.}
\label{fig1}
\end{figure}

Let us use the genetic algorithm described in section \ref{GA} to exemplify the $q$-exponential fitting method employed. Disregarding elitism and taking a generation limit of $t=100$ generations, we can perform $R=20$ simulations\footnote{Generating 20 genealogical networks of 10000 individuals (100 individuals per generation in 100 generations).} and calculate the ETV values for each individual. Now, we calculate the frequency of each ETV value using Equation \ref{freq} and create the corresponding probability density distributions. For visualization purposes, we apply the $q$-logarithm function, defined as $ln_q(x) \equiv [x^{(1-q)} - 1]/(1- q)$, with $ln_1x \equiv ln(x)$, in the calculated probabilities. If the distribution fits a $q$-exponential function, then there is a pair of values $(q,x_0)$ in which the data will be adjusted on a straight line \cite{picoli2009q}. Applying the $ln_q(x)$ function on both sides of Equation \ref{eq1}, we obtain
\begin{equation}
ln_q \;p_{qe}(x)=ln_q\; p_0 -[1+(1-q)ln_q\; p_0]\frac{x}{x_0}.
\end{equation}
Since $p_0=(2-q)/x_0$ (normalization condition, see Equation \ref{eq1}), determining the values of $q$ and $x_0$ will now consist of verifying at which point the most number of higher frequency ETVs best tend to concentrate on a straight line. Figure \ref{fig1} a) presents the resulting PETV distribution (average of 20 simulations) and the best-fitting $ln_{q}\; p_{qe}$ curve for $q=1.19$, $p_0=1.0421$, and $x_0=0.7773$.

The traditional method for illustrating the power law is, however, through log-log scale plots in which straight asymptotes are drawn on to illustrate the power law. We will therefore adopt the log-log scale representation in all of our next plots, making it so that our results can be compared to those divulged in other similar articles in the literature. Like this, in Figure \ref{fig1} b), where we now use log-log scale instead of $q$-logarithm scale, note that there is a concavity in the distribution and that the $q$-exponential distribution adjusts the data reasonably well in all ETV values.

This process can be reproduced for several other configurations, using or disusing several other mechanisms, allowing us to study the resulting power law dynamics and how each employed mechanism influences it.

\section{Results}
In this section, we exhibit the power law dynamics observed in the genealogical networks generated with the genetic algorithm described in section \ref{GA}. The conditions for constructing the genealogical networks, PETV distributions, and corresponding best-fitting $p_{qe}$ curves are the same as the example in section \ref{Dfm}.

Using as reference the data available in TSPLIB \cite{reinelt1991tsplib}, we ran the GA on several TSP problems (varying from 14 to 42 points), generated the correspondent PETV distributions, and analyzed the resulting power law dynamics over time. We obtained very similar results for all problems tested\footnote{As expected, since genealogical networks have been found be little sensitive to the ﬁtness landscape on which evolution occurs \cite{whitacre2009making}.}. Here, we will present the results for the most complex problem tested, swiss42 (42 cities Switzerland), divided into two parts. The first part shows the results obtained while not using elitism. The second part shows the results obtained while using elitism and other mechanisms. Finally, we compare the power law dynamics observed in both cases and discuss the influence of elitism and other mechanisms.

\subsection{First experiment}
\label{SemEl}
\begin{figure}[!ht]
	\centering
	\centerline{\includegraphics[scale=0.55]{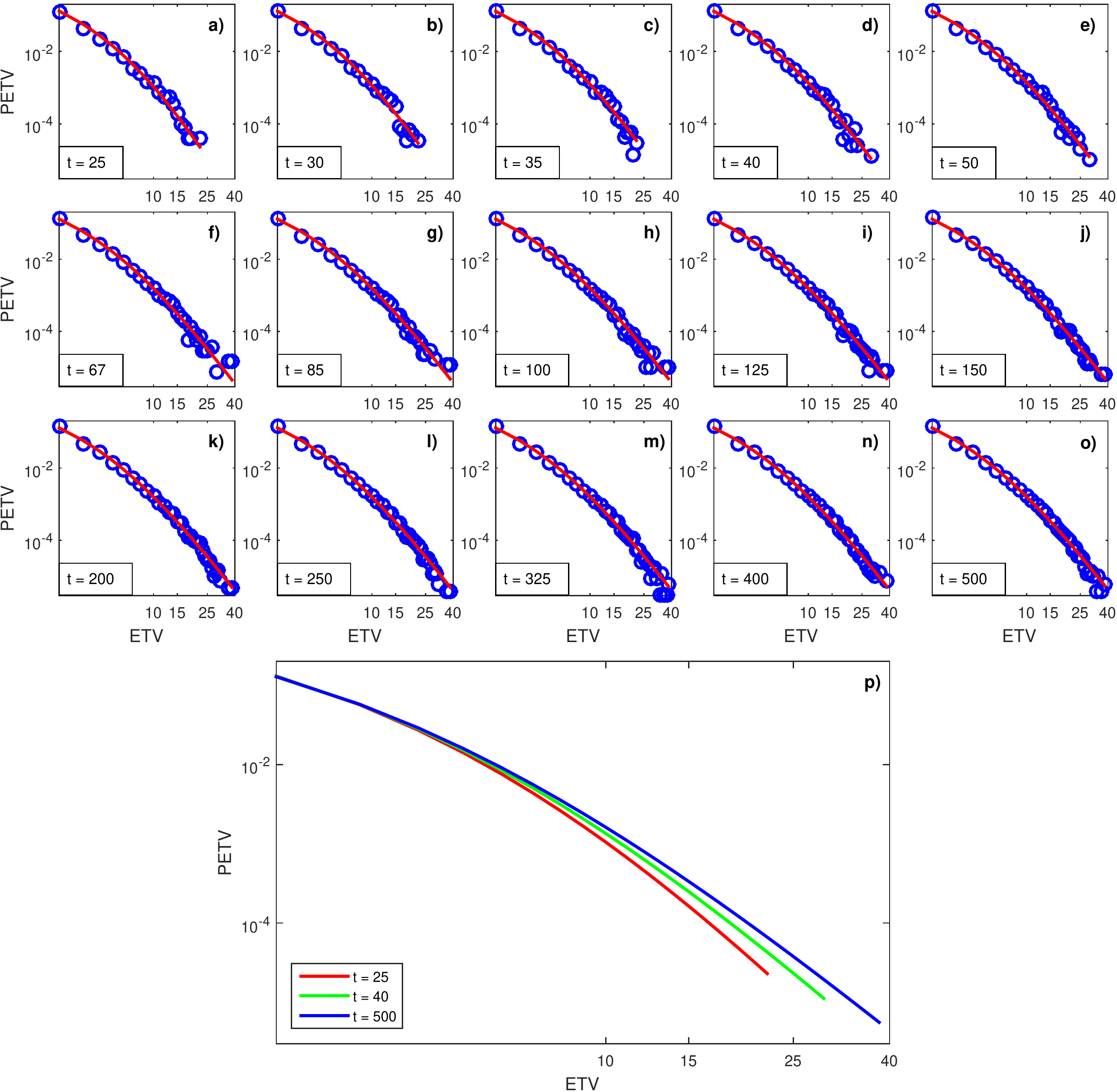}}
		\caption{PETV vs. ETV distributions (blue markers) and best-fitting $p_{qe}$ curves (red curves) for a) $t = 25$, b) $t = 30$, c) $t = 35$, d) $t = 40$, e) $t = 50$, f) $t = 67$, g) $t = 85$, h) $t = 100$, i) $t = 125$, j) $t = 150$, k) $t = 200$, l) $t = 250$, m) $t = 335$, n) $t = 400$, and o) $t = 500$ generations. p) Best-fitting $p_{qe}$ curves for $t = 25$, $t = 40$, and $t = 500$ generations. Note that, over time, there is a progressive increase in the frequency of occurrence of higher ETV values. Also note that the distribution becomes less concave over the generations. From generation 500 on, variations in the shape of the distribution are almost imperceptible.}
	\label{fig2}
\end{figure}

In our first experiment, we disregarded elitism and ran the GA up to $t = 500$ generations. After 20 runs (thus generating 20 genealogical graphs of 50000 individuals each), we analyzed the impact of all individuals in all populations across all generations. We calculated the ETV occurrence probability values (PETV) and plotted the resulting probability density distributions in fifteen generation intervals. Finally, we determined the best-fitting $q$-exponential curves for each interval and plotted these over the probability density distributions. The results can be seen in Figure \ref{fig2}, where we exhibit the obtained PETV vs. ETV distributions and its respective best-fitting $p_{qe}$ curves.

In analyzing Figure \ref{fig2}, first observe that the behavior pattern of these distributions can be satisfactorily represented by a family of $q$-exponential curves (where a good fit can be made with the appropriate $q$, $p_0$, and $x_0$ values) at any generation. Also note that, without elitism, a slow evolution process occurs. This is evidenced by the fact that, after 500 generations, the maximum ETV value achieved was only 38 (on a scale of 1 to 100).

As for the distribution dynamics, note that there is a progressive increase in the frequency of occurrence of higher and higher ETV values as generations pass. This leads to an increase in the number of points of the distribution and to the lifting of its tail (see Figure \ref{fig2} p)) due to the increasingly larger PETV values in higher ETV positions. This process makes the distribution less and less concave (or more and more straight) over time. This ``straightening'' speed is more pronounced in the first generations and decreases in intensity (and tends to saturate) as the population evolves.

\begin{figure}[!ht]
	\centering
	\includegraphics[scale=0.5]{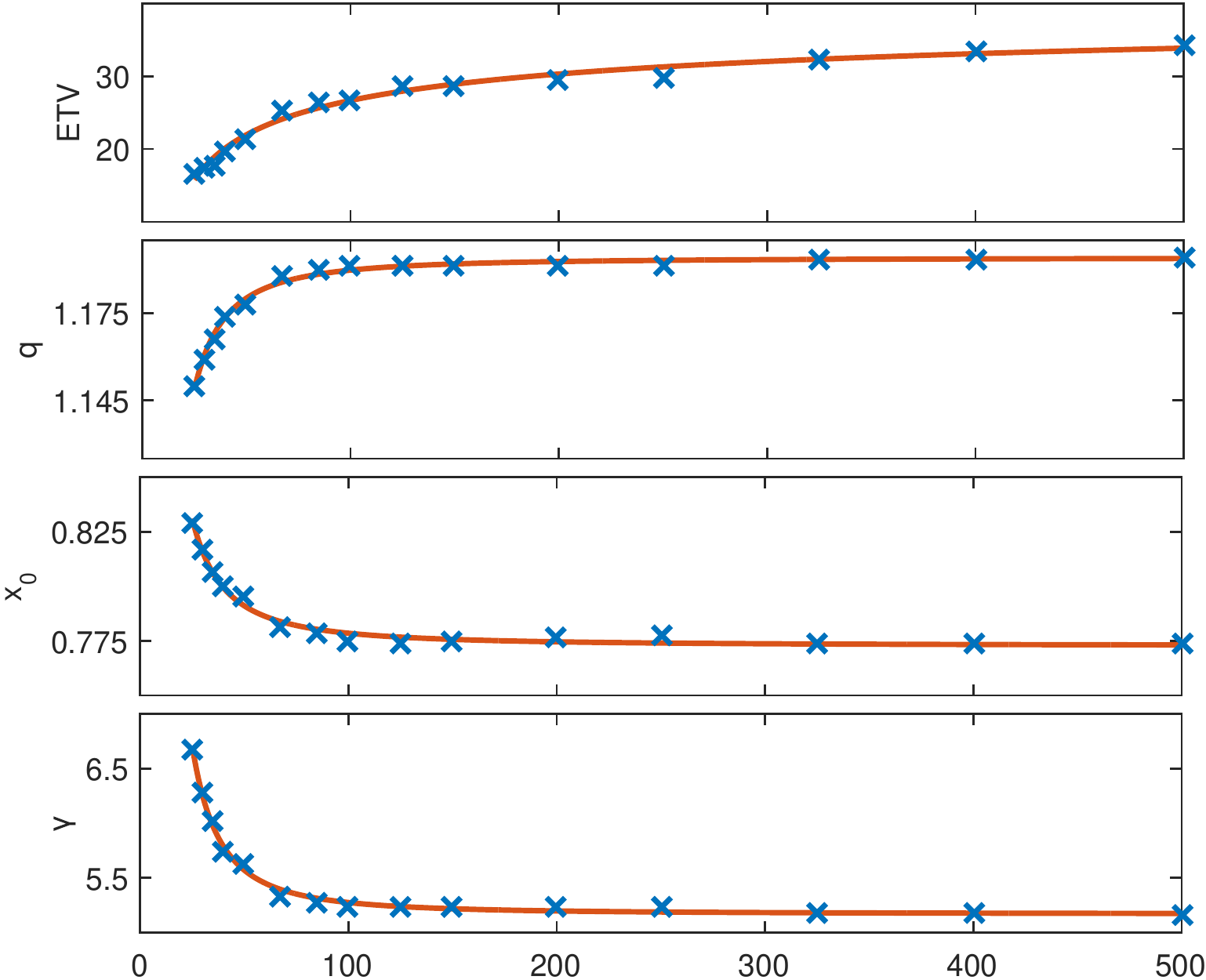}
		\caption{Evolution of the maximum average ETV values, calculated $p_{qe}$ parameters, $q$ and $x_0$, and corresponding power law scaling factors, $\gamma$, over the generations. \textbf{$\times$ markers:} values obtained for 25, 30, 35, 40, 50, 67, 85, 100, 125, 150, 200, 250, 335, 400 and 500 generations. \textbf{Red curves:} best-fitting interpolating curves.}	
	\label{fig3}
\end{figure}

 The power law dynamics require that the values of the $q$, $x_0$, and $p_0$ parameters be adjusted at each generation. In Figure \ref{fig3}, we exhibit the evolution of the maximum average ETV values, the variation of the $q$ and $x_0$ parameters on the $p_{qe}$ curves, and the corresponding power law scaling factors, $\gamma$, over the generations\footnote{$p_0=(2-q)/x_0$ due to normalization. The asymptotic form of the $q$-exponential is $x^{-\gamma}$, where $\gamma=1/(q-1)$.}. Note that there is a regularity in the observed values ($\times$ markers), which allows us to fit these into $f(t)=at^b+c$ functions, where $t$ is the generation value, and $a$, $b$, and $c$ are the interpolation parameters. Also note that the $q$ curve is positively correlated to the maximum average ETV curve, whereas the $x_0$ and $\gamma$ curves are negatively correlated to the maximum average ETV and $q$ curves. The best-fitting parameters and regression values for the interpolating curves in Figure \ref{fig3} are shown in table \ref{tab1}.

\begin{table}[!ht]
	\centering
			\caption{$a$, $b$, $c$ parameters and regression, R, for the best-fitting $ETV$, $x_0$, $q$, and $\gamma$ interpolating curves obtained. Fitting function: $f(t)=at^b+c$.}
	\begin{tabular}{|c|c|c|c|c|}\hline
		    Parameters  &   $a$   &   $b$ &  $c$  & R \\ \hline
		    ETV   & -74.2 &  -0.2435 & 50.35 & 0.983 \\ \hline
		   $x_0$  &  8.038 & -1.525  & 0.7696 & 0.984 \\ \hline
		    $q$   &  -6.306 & -1.527 & 1.196 & 0.988 \\ \hline
	   $\gamma$   &  795.3 & -1.941  & 5.169 & 0.992 \\ \hline
	\end{tabular}
	\label{tab1}
\end{table}

After analyzing the variation ratios shown in Figure \ref{fig3}, we realize that the large variations in the $q$ and $x_0$ parameters (between generations $t=25$ and $t=150$) are due to a decreasingly rapid emergence of individuals with higher ETV values in this interval. As the emergence of higher impact individuals decreases (generation 150 onwards), a regime is reached, and changes in the distribution shape occur very slowly (with a maximum average ETV variation of only 4 points between generations $t=250$ and $t=500$). Comparing these results to the $p_{qe}$ curves in Figure \ref{fig2} shows us that intervals of higher variation ratios (earlier generations) correspond to the most concave $p_{qe}$ curves, whereas the intervals of lower variation ratios (later generations) correspond to the least concave (or straighter) $p_{qe}$ curves. For every simulation, the power law dynamics observed using the $q$-exponential curves is that of initially concave lines that become more and more straight over time. This ``straightening’’ speed is higher in the first generations but tends to decrease in intensity and gradually stagnate.

\subsection{Second experiment}
Aging and maximum edge limitation provide similar results in causing power law deviating behaviors \cite{albert2002statistical}. Since, in our case, maximum edge limitation mostly only impacts the number of points of the distribution, we will only show the results obtained with the aging mechanism.

\begin{figure}[!ht]
	\centering
	\includegraphics[scale=0.55]{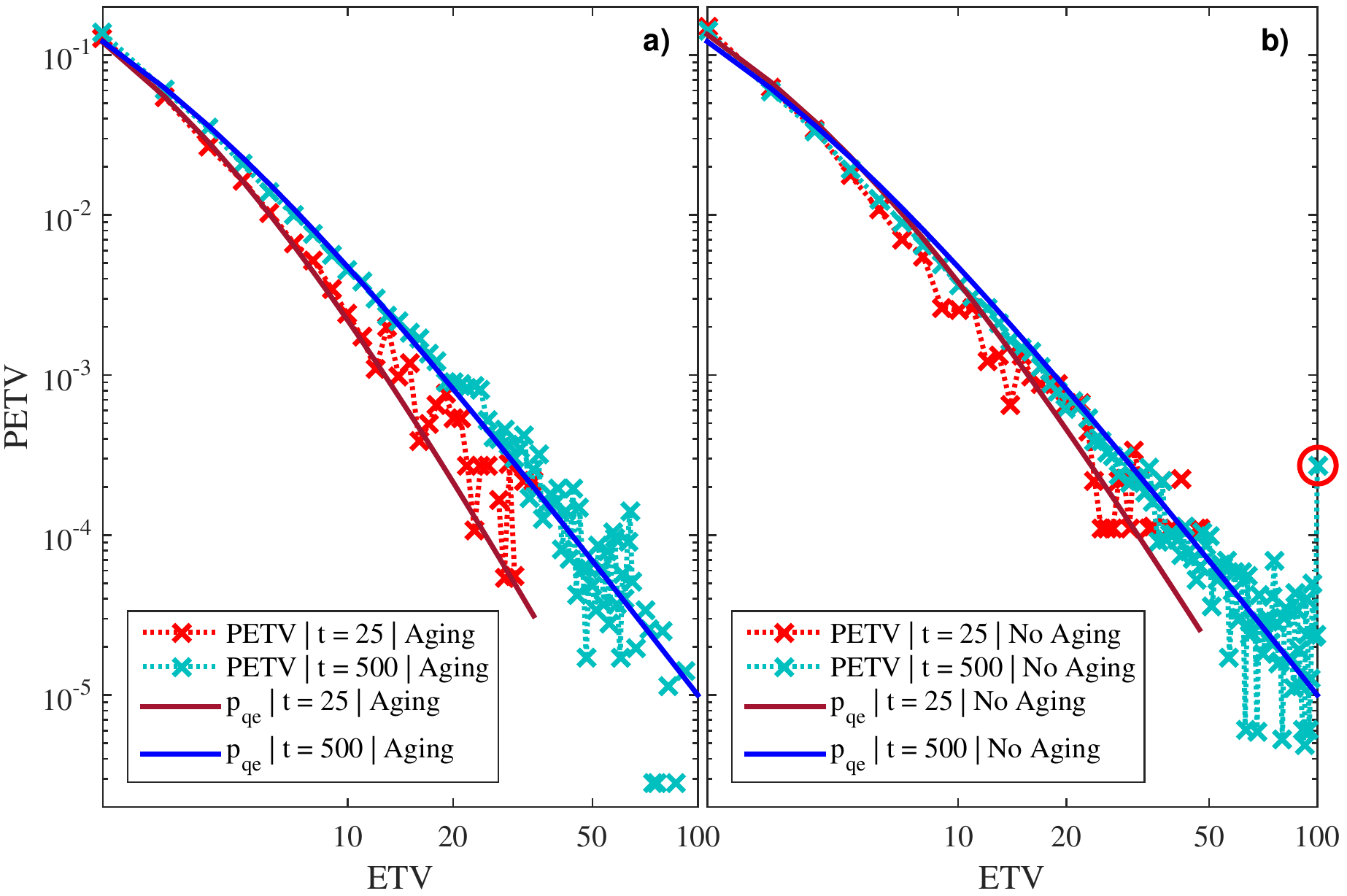}
		\caption{
		\textbf{$\times$ markers:} mean values of PETV vs. ETV at $t = 25$ and $t = 500$ generations. \textbf{Red curves:} best-fitting $p_{qe}$ curves at $t = 25$ and $t = 500$ generations. Average of 20 simulations.
		a) Use of elitism and fast-aging ($m = 2$). At $t = 25$ generations, the calculated $p_{qe}$ parameters are $q = 1.25$ and $x_0 = 0.665$, and the corresponding power law scaling factor is $\gamma = 4$. At $t = 500$ generations, the calculated $p_{qe}$ parameters are $q = 1.35$ and $x_0 = 0.635$, and the corresponding power law scaling factor is $\gamma = 2.8571$. 
		b) Use of elitism only. At $t = 25$ generations, the calculated $p_{qe}$ parameters are $q = 1.26$ and $x_0 = 0.795$, and the corresponding power law scaling factor is $\gamma = 3.8462$. At $t = 500$ generations, the calculated $p_{qe}$ parameters are $q = 1.345$ and $x_0 = 0.625$, and the corresponding power law scaling factor is $\gamma = 2.8986$.
		}
	\label{fig4}
\end{figure}

For our second experiment, we employ the same configurations used in the first experiment, with the sole exception of the use of elitism and/or the aging mechanism. The results can be seen in Figure \ref{fig4}, where we exhibit the obtained PETV vs. ETV distributions for $t=25$ and $t=500$ generations, and its respective best-fitting $p_{qe}$ curves. The remaining curves, including the $t=40$ shown in the first experiment (Figure \ref{fig2} p)), were omitted here to not overly ``pollute'' the figure, but they also presented very similar fits. The evolution of the maximum average ETV values, the variation of the $q$ and $x_0$ parameters on the $p_{qe}$ curves, and the corresponding power law scaling factors, $\gamma$, over the generations were similar to those shown in Figure \ref{fig3}.

\begin{figure}[!ht]
	\centering
	\includegraphics[scale=0.5]{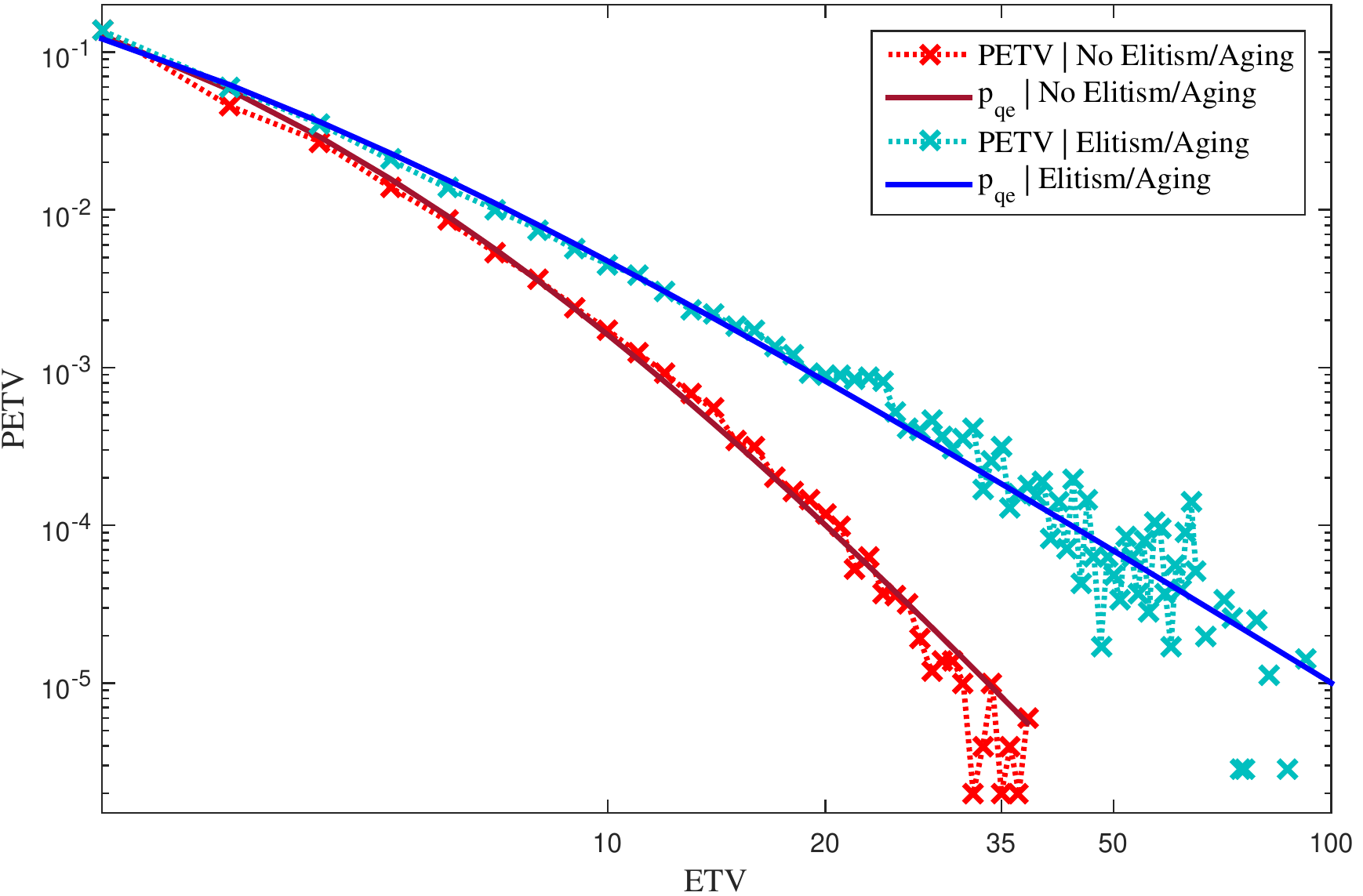}
		\caption{Comparison between the PETV vs. ETV distributions ($\times$ markers) and corresponding best-fitting $p_{qe}$ curves at $t = 500$ generations for both experiments. Fitting parameters and corresponding power law scaling factors: $q = 1.194$, $x_0 = 0.773$, $p_0 = 1.042$, $\gamma = 5.1546$ (dark red curve) and $q = 1.350$, $x_0 = 0.635$, $p_0 = 1.040$, $\gamma = 2.8571$ (blue curve).}
	\label{fig5}
\end{figure}

In analyzing Figure \ref{fig4}, first note that the behavior pattern of these distributions can, once again, be satisfactorily represented by $q$-exponential curves. Second, note that, in both cases, maximum ETV value is achieved at $t=500$ generations. A closer analysis will show that maximum ETV value was achieved as soon as generation $t=250$ with or without aging. This result shows us that the use of elitism, by itself, caused an accelerated emergence of the power law (that is, a rapid ``straightening" in the $q$-exponential curve). In fact, elitism sped up this ``straightening" so much that, compared to the curves in the first experiment, generation $t = 15$ with elitism is already as ``straight" as generation $t = 250$ without elitism. This difference increases with the number of generations until it reaches a saturation limit for both. A visual representation of how much elitism affected the power law dynamics here is shown in Figure \ref{fig5}, where we plot the PETV vs. ETV distributions and corresponding best-fitting $p_{qe}$ curves at generation $t = 500$ for both experiments.

A comparative analysis on the graphs in Figure \ref{fig4} shows that the use of aging caused an initial difference in ``straightness'' in the earliest generations, meaning that power law emerged faster without aging. This difference diminishes along the generations, and the two distributions become similar after saturation. In fact, at $t = 500$ generations, both best-fitting $p_{qe}$ curves are almost identical. The main difference between both graphs is the appearance of several ``strange points" at the tail of the distribution (highlighted by the red circle), representing a paradoxical high probability of appearance of individuals of the highest impact. This anomaly also appears in most of the results presented in \cite{whitacre2009making}. A closer analysis of the populations across all generations shows that this anomaly appears as soon as fitness balance\footnote{Fitness balance is the state in which all individuals in the population have the same fitness value.} is achieved. At this point, elitism ``kills'' any new offspring, genetic variability is lost in the population, and the evolution process is finished with only clones being added to the population. This result is an indication that the use of elitism can lead to misleading insights unless other restrictive mechanisms (such as fast-aging) are employed, or the simulation is stopped as soon as genetic variability is lost (which can be tricky, since fitness plateaus, followed by fast evolution, are sometimes common on more complex problems).

Finally, from Figures \ref{fig4} and \ref{fig5}, note the difference in the values of the corresponding power law scaling factors in both experiments. Without elitism, $\gamma = 5.1546$ at $t = 500$ generations. With elitism, $\gamma = 2.8571$ (with aging) or $\gamma = 2.8986$ (without aging), at $t = 500$ generations. The scaling factors obtained using elitism are now in the typical range of $2 \leq \gamma \leq 3$ \footnote{Save the occasional exception, the power law scaling factor typically lies in the range of $2 \leq \gamma \leq 3$ \cite{clauset2009power}.}. This result is further indication that the use of elitism can significantly influence the power law dynamics in population-based optimization algorithms.

\section{Discussion}
\subsection{Similarity to research based on empirical data}
Several papers in the literature present evidence that real complex systems follow a $q$-exponential power law \cite{malacarne2001q, yamada2008q, takahashi2011depressive, takahashi2008psychophysics, cajueiro2006note, picoli2005statistical, briggs2007modelling, politi2008fitting, jiang2008scaling, kaizoji2006interacting, kaizoji2004inflation, anastasiadis2009characterization, tsallis2000citations, oikonomou2008nonextensive}. In our experiments, we implemented several of the conditions found in real complex networks, such as unclear
or multiple optimal solutions, preferential attachment, mutation, reproduction by gene recombination, and aging \cite{barabasi1999emergence, amaral2000classes, albert2002statistical}. We also incorporated several of the mechanisms reported to cause significant power law deviations, such as the severe mutation, insertion of historically uncoupled individuals, maximum edge limitation, and fast aging \cite{albert2002statistical}. Ultimately, we verified that the observed individual impact value probability distributions could be satisfactorily modeled, at any point, by $q$-exponential curves. These $q$-exponential curves presented concavities very similar to those highlighted in \cite{whitacre2009making, albert2002statistical} (see Figure \ref{fig4} a)) and in several other papers based on empirical data \cite{malacarne2001q, yamada2008q, takahashi2011depressive, takahashi2008psychophysics, cajueiro2006note, picoli2005statistical, briggs2007modelling, politi2008fitting, jiang2008scaling, kaizoji2006interacting, kaizoji2004inflation, anastasiadis2009characterization, tsallis2000citations, oikonomou2008nonextensive, redner1998popular, seglen1992skewness, solow2003testing, clauset2009power} (see Figure \ref{fig2} and Figure \ref{fig4} b)).

Elitism, as applied in evolutionary algorithms, rarely occurs in the real world. In our numerical experiments, we found that, without elitism, populations evolve more slowly and randomly, but genetic variability is generally higher (more diverse populations). We also found this to be the situation in which we achieved our best $q$-exponential data fittings at any stage of evolution. The maintenance of this distribution shape from early generations to a nearly stationary regime (population stabilization) resembles many results observed in empirical data from real population structures \cite{malacarne2001q, yamada2008q, takahashi2011depressive, takahashi2008psychophysics, cajueiro2006note, picoli2005statistical, briggs2007modelling, politi2008fitting, jiang2008scaling, kaizoji2006interacting, kaizoji2004inflation, anastasiadis2009characterization, tsallis2000citations, oikonomou2008nonextensive, redner1998popular, seglen1992skewness, solow2003testing, clauset2009power}.

\subsection{Power law dynamics}
The emergence of a power law distribution implies that most individuals have a negligible impact on population dynamics and do not provide useful information (i.e., act as noise). In several research papers, particularly in  \cite{whitacre2009making, albert2002statistical}, the authors analyze the behavior of empirical and/or artificial data distributions under various configurations and observe the emergence and ``breakdown" of power law behaviors under one condition or another. In general, at the end of these analyzes, the authors conclude that most systems analyzed initially do not have population dynamics defined by a power law, but over the generations, or under specific conditions, the system evolves to reach that state.

In our work, we obtain very similar results, but we analyze them from a different perspective. Instead of using a conventional power law curve, we used Tsallis's $q$-exponential distribution (a generalization of the power law) and noticed that cases previously understood as ``power law deviations'' fit perfectly into a family of $q$-exponential curves. We found that $q$-exponential functions could satisfactorily model the power law dynamics throughout the entire process (from the first to the last generations) by simply adjusting its $q$ and $x_0$ parameters. Both these parameters and its corresponding power law scaling factors, $\gamma$, are time-dependent and follow a $f(t)=at^b+c$ pattern, where $t$ is the number of generations. These results provide further evidence that nonextensive statistics and scale-free networks are intimately related (connection  conjectured in \cite{tsallis2004should} and discussed in \cite{soares2005preferential}). They also imply that power law behavior stands true regardless of many previously reported ``power law deviating'' mechanisms employed, but its shape is time-dependent.

\section{Conclusion}
\label{Conc}
Power law is a statistical phenomenon found in many population samples. Power law behavior with several different shapes can also be observed in genealogical networks. In this paper, we investigated the dynamics and relation to population evolution over time of power law behaving probability distributions observed in numerically generated genealogical networks. First, we used a genetic algorithm to generate several genealogical graphs and measured the impact of all individuals in all populations across all generations. Then, we analyzed the resulting individual impact value probability distributions and studied the dynamics of evolution of said populations through nonextensive statistics. Like this, we verified that the emergence of power law in these distributions has a dynamic behavior over time. This dynamic development can be well described by a family of $q$-exponential distributions in which the $q$ and $x_0$ parameters follow a $f(t)=at^b+c$ pattern. Finally, we discussed the use of elitism and other restrictive mechanisms and their effects on the power law dynamics. We found that elitism, while a valuable tool that boosts conversion speed, also dramatically influences the observed power law dynamics, sometimes even creating strange anomalies that could lead to misleading insights.

The results presented in this paper are important and have many implications. First, they indicate that the causes responsible for the formation of the power law are also dynamically changing within the genealogical network. Second, they provide evidence that power law behavior stands true regardless of many previously reported ``power law deviating'' mechanisms employed. Third, they show that the different power law shapes (with several scaling factors) and deviations observed in our genealogical networks are, in fact, static images of a time-dependent dynamic development that can be described using $q$-exponential distributions. Fourth, they imply that the different power law shapes and deviations reported in similar papers may also be static images of a time-dependent dynamic development. Fifth, they show that elitism (commonly used as default in population-based optimization algorithms) significantly influences the power law scaling factors and deviations observed. Finally, they provide further evidence for the conjecture that relates nonextensive statistical mechanics with complex networks.

\section*{Declaration of competing interest}
The authors declare that they have no known competing financial interests or personal relationships that could have appeared to influence the work reported in this paper.

\section*{Data availability}
Data underlying the results presented in this paper are not publicly available at this time but may be obtained from the authors upon reasonable request.

\bibliographystyle{IEEEtran}

\bibliography{ref}

\end{document}